\documentclass[a4paper]{jpconf}
\usepackage{graphicx}
\usepackage{mine}

\RequirePackage[capitalise]{cleveref}

\begin{document}
\title{Theory prediction in PDF fitting}

\author{Andrea Barontini\textsuperscript{1}, Alessandro Candido\textsuperscript{1}, Juan M. Cruz-Martinez\textsuperscript{2},
Felix Hekhorn\textsuperscript{1}, Christopher Schwan\textsuperscript{3}}
\address{%
  \textsuperscript{1} TIF Lab, Dipartimento di Fisica, Universit\`a degli Studi di Milano and INFN
  Sezione di Milano, Via Celoria 16, 20133, Milano, Italy\\
  \textsuperscript{2} CERN, Theoretical Physics Department, CH-1211 Geneva 23, Switzerland\\
  \textsuperscript{3} Universit\"at W\"urzburg, Institut f\"ur Theoretische Physik und
  Astrophysik, 97074 W\"urzburg, Germany
}

\begin{abstract}
Continuously comparing theory predictions to experimental data is a common task in analysis of particle physics such as
fitting parton distribution functions (PDFs). However, typically, both the computation of scattering amplitudes and the
evolution of candidate PDFs from the fitting scale to the process scale are non-trivial, computing intesive tasks.
We develop a new stack of software tools that aim to facilitate the theory predictions by computing FastKernel (FK)
tables that reduce the theory computation to a linear algebra operation. Specifically, I present PineAPPL, our workhorse
for grid operations, EKO, a new DGLAP solver, and yadism, a new DIS library. Alongside, I review several projects that
become available with the new tools.
\end{abstract}

\section{Introduction}

Parton distribution functions~\cite{Amoroso:2022eow} (PDFs) describe the dynamics of the elementary partons, such as quarks and gluon,
inside hadrons, such as the proton. PDFs are defined in factorization theorems in high-energy scattering
and they encode the non-perturbative physics reigning inside hadrons.
The extraction of PDFs relies on three pillars: first, the precise measurements of observables at experiments, such as the LHC,
second, the reliable theoretical predictions of the associated observable, and, third, combining the former two inside
a fitting framework. While either of these three steps poses several computational challenges,
we discuss here only the efficient computation of the theory predictions.

The computation of the various high-energy observables that enter a PDF fit are pursuit by different groups
applying a range of strategies using several dedicated programs tailored to the specific case at hand.
Yet, in a PDF fit one wants to include \textit{consistently} all predictions available, which can be up to
4500 data points across almost 100 different datasets (as is the case, e.g., in \cite{Ball:2021leu}).

To achieve this goal we develop a framework of software tools, dubbed \pineline~\cite{pineline}, that provides a
easy accessible way to include theory prediction in QCD fitting applications such as PDF fits.
We put an explicit emphasis on the scalability of the procedure to ensure future measurements
form new or existing experiments~\cite{Gao:2017yyd,Accardi:2012qut,Anderle:2021wcy} can be included
seamlessly. We also strive to track all runcards and meta data in the generated objects to ensure,
we are able to reproduce the existing results. Finally, we stress that all participating
programs are developed Open Source to facilitate the interaction with users and
developers (following the efforts of the NNPDF fitting code~\cite{NNPDF:2021uiq}).

\section{Technical Overview of the \pineline{}}
\label{sec:tech}
We refrain here from repeating all technical details from the previous
publication~\cite{pineline}, but give in the following just a
brief overview of the framework and its the participating programs.

\begin{figure}
  \centering
  \includegraphics[width=\textwidth]{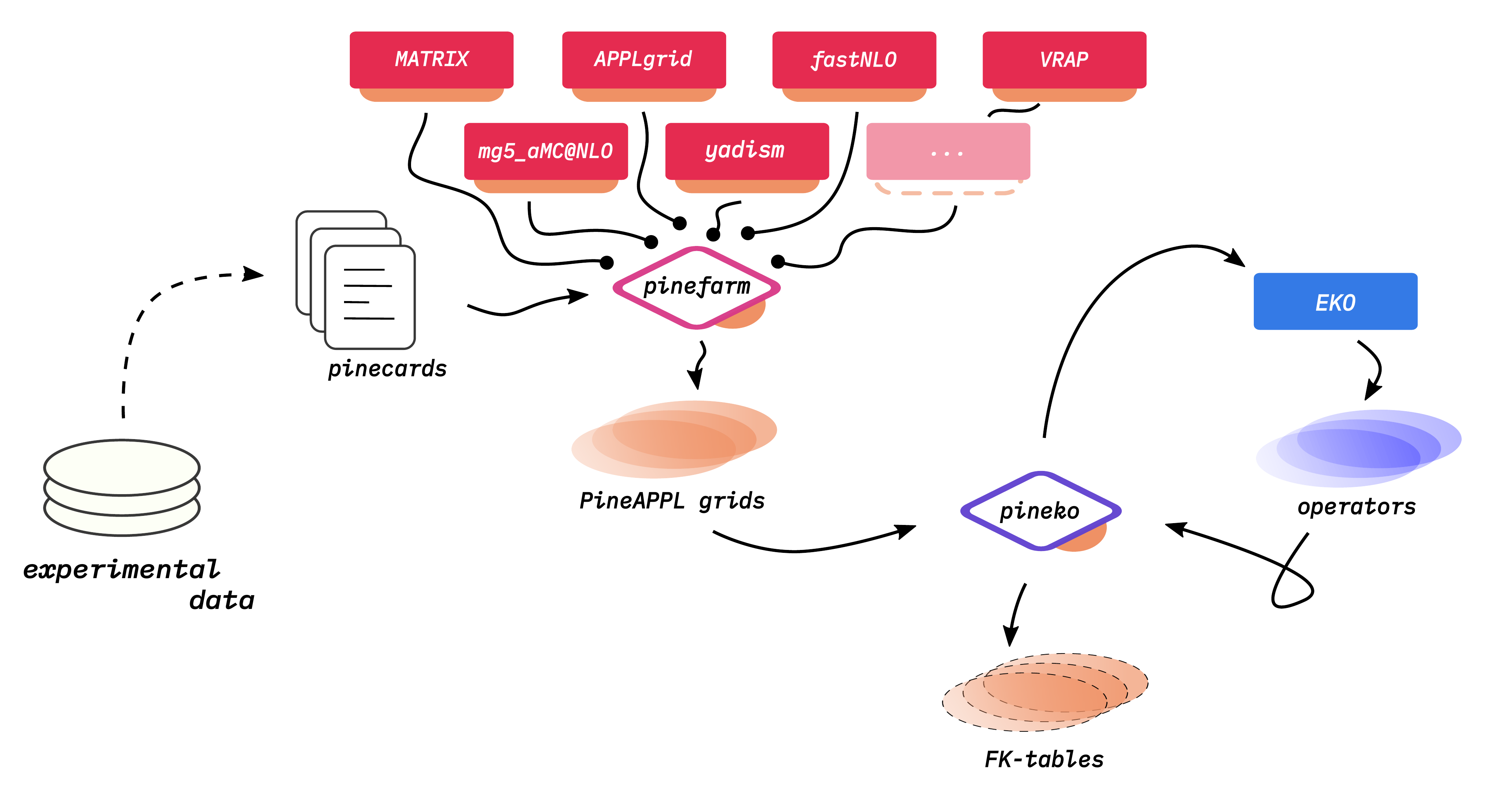}
  \caption{\label{fig:flow}
  Flow diagram showing the overall pipeline architecture and deliverables in the case of parameter fits. Arrows in the picture indicate
  the flow of information (together with the execution order) and the orange insets on other elements indicate an interface to PineAPPL. The
  programs \pinefarm{} and \pineko{} act as interfaces between other programs and the deliverable objects, represented by ovals. These objects can
  be PineAPPL grids (orange) or Evolution Kernel Operators (blue).
  }
\end{figure}

The core part of the framework is to use PDF independent interpolation grids, as provided by
\texttt{PineAPPL}~\cite{Carrazza:2020gss,christopher_schwan_2023_7499507} to be able to reuse
predictions for any candidate PDF. We then extend the idea to the concept of FastKernel (FK) tables~\cite{NNPDF:2014otw}
where we now include also the perturbative evolution of PDFs inside the interpolation grid.
FK tables can then be used efficiently inside a PDF fit as the usually complicated
convolution between candidate PDF and object is now replaced by a simple linear algebra
operation. To accomplish the various steps we develop the following programs:
\begin{itemize}
  \item \texttt{pinefarm} acts as an interface to existing Monte Carlo generators to produce \texttt{.pineappl} grids
    containing the partonic information
  \item \texttt{yadism} provides structure functions in deep-inelastic scattering~\cite{candido_alessandro_2022_6285149} (DIS) (and is interfaced to \pinefarm)
  \item \texttt{eko} provides the solution to the (perturbative) evolution equations in term of evolution kernel operators~\cite{candido_alessandro_2022_6340153,Candido:2022tld}
  \item \texttt{pineko} combines partonic grids and evolution kernel operator into an FK table
\end{itemize}
The flow of programs is summarized in \cref{fig:flow}.

\section{Benchmarking}

The major focus of the \pineline is to join several dedicated programs into a consistent framework for
theory predictions. However, it is relying on those programs to provide the necessary physical ingredients
as the framework itself is (almost) physics agnostic and can indeed not only be applied to PDF fitting,
but also the determination of other factorized objects, such as fragmentation functions~\cite{AbdulKhalek:2022laj},
or beyond standard model (BSM) searches~\cite{Ball:2022qtp}.

Still, it is convenient to develop some new programs which are tailored for the specific needs and concepts
of the \pineline. Specifically, we develop \eko~\cite{Candido:2022tld,candido_alessandro_2022_6340153}, a new \dglap
solver~\cite{Dokshitzer:1977sg,Gribov:1972ri,Altarelli:1977zs}, and \yadism~\cite{candido_alessandro_2022_6285149},
a new provider of deep-inelastic structure functions. Either of these tasks is an already solved problem
for which several implementation exist~\cite{Bertone:2013vaa,Botje:2010ay}, yet none of them provide the
exact deliverable that we require here.

In order to benchmark \eko and \yadism to former implementations we develop \banana~\cite{barontini_andrea_2023_7636142}.
\banana provides a convenient benchmarking framework that turned out very useful during the development of both codes.
It features a full-fledged database system (based on \texttt{SQLite}) to map input runcards to the respective outputs
of both programs, the one to be benchmarked and the reference implementation. This allows a seamless comparison between the
ongoing development and the reference implementation, such as investigating the difference in certain regions of the
parameter space (which often can be mapped back onto a certain region in physics space). Since the reference implementation
is fixed we implement a caching algorithm which ensures a fast iteration in the development process. Keeping a history
of the runs allows us also to easily compare our program in two different version which again simplifies the development
significantly.

In order to facilitate the access to the database we provide an interface built on top of \texttt{IPython}~\cite{PER-GRA:2007},
dubbed \texttt{navigator}, that allows to easily query and retrieve records from the database.
Moreover, being embedded into a live Python interpreter makes the \texttt{navigator} flexible and powerful.

We also add a few simple operations to the \texttt{navigator} such as the difference between program outputs. This was very helpful e.g.\ in the implementation
of the FONLL prescription~\cite{Forte:2010ta} for DIS structure function, which can be boiled down to the line
\begin{equation}
  F^{\rm FONLL} = F^{(n_l+1)} + F^{(n_l)} - F^{(n_l,0)}
\end{equation}
where $F$ refers to any DIS structure function and the various expressions on the right hand side refer to specific
physical prescriptions. Each of these prescriptions are well-defined on their own, but already non-trivial, composite objects.
Thus, having the possibility to benchmark them one at a time and subtracting either from the combination
is very beneficial.

\begin{figure}
  \centering
  \includegraphics[width=\textwidth]{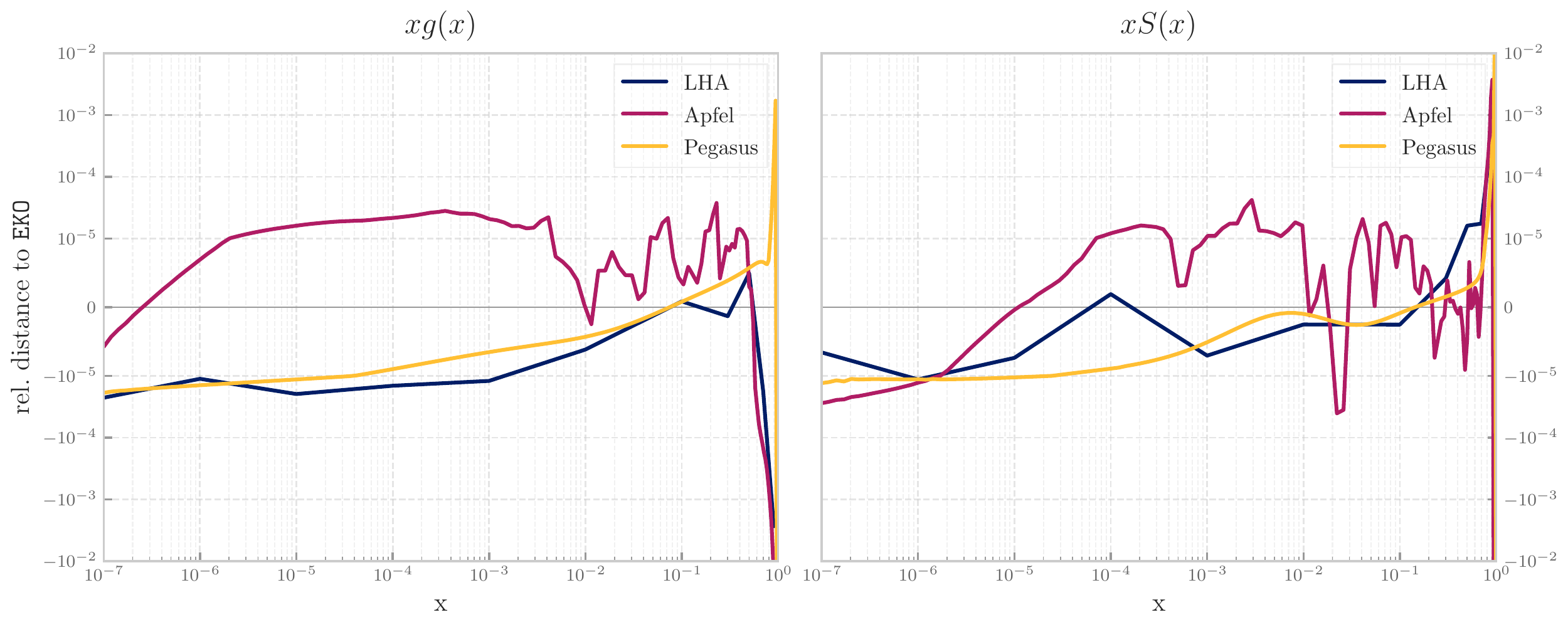}
  \caption{\label{fig:LHA}
  Part of \cite[Fig.~1]{Candido:2022tld} comparing different implementation of the evolution
  equations to the LHA benchmark tables~\cite{Giele:2002hx,Dittmar:2005ed} using \texttt{banana}.
  On the left (right) panel the gluon distribution $g(x)$ (singlet distribution $S(x)$) is plotted as
  a function of the momentum fraction $x$.
  }
\end{figure}

While \banana provides the necessary framework that abstracts most tasks away, \eko and \yadism have to implement each
a specialization of the framework, dubbed \texttt{ekomark} and \texttt{yadmark} respectively, as indeed they compute
different, but related objects. These specialized programs are developed in unison together with the main programs.
\texttt{ekomark} currently provides interfaces to the LHA benchmark tables~\cite{Giele:2002hx,Dittmar:2005ed},
\texttt{PEGASUS}~\cite{Vogt:2004ns}, \apfel~\cite{Bertone:2013vaa}, and to
any LHAPDF~\cite{Buckley:2014ana} set. \texttt{yadmark} currently provides interfaces to \apfel~\cite{Bertone:2013vaa},
\texttt{QCDNUM}~\cite{Botje:2010ay}, and \texttt{xspace\_bench}~\cite{Forte:2010ta}.
In \cref{fig:LHA} we show an application of the \texttt{ekomark} benchmarks using the
\banana~framework.

\section{Conclusion and outlook}
With the \pineline we aim to provide an simple framework to generate theory predictions
from a common input and to deliver a unified output. This can be beneficial for the fitting
of parton distribution functions~\cite{Amoroso:2022eow} (PDF) or any factorized function such as fragmentation
functions~\cite{AbdulKhalek:2022laj}. We split the necessary tasks to produce FastKernel tables~\cite{NNPDF:2014otw}
into separate programs, namely \pinefarm, \eko, and \pineko, each focusing only on the specific part.
This allows to extend the framework to more involved physics cases such as the determination of the
photon PDF~\cite{Xie:2021equ,Cridge:2021pxm,Bertone:2017bme}, the inclusion of missing higher order uncertainties into
PDF fits \cite{NNPDF:2019ubu}, or the extension to next-to-next-to-next-to-leading order (N3LO) perturbation
theory~\cite{McGowan:2022nag,Caola:2022ayt}.

While developing the \pineline, we also develop two participating programs: \eko and \yadism.
Either one is reimplementing algorithms and ingredients from already known calculations.
It is thus imperative to ensure the new code reproduce previous implementations in a
benchmarking process. To achieve this task we also develop \banana which provides
a sophisticated benchmarking framework tailored for the case at hand which allows
a reliable and repeated execution of benchmarks.

\section*{Acknowledgments}
A.C.\ and F.H.\ are supported by the European Research Council under the
European Union's Horizon 2020 research and innovation Programme (grant
agreement number 740006).
C.S.\ is supported by the German Research Foundation (DFG) under reference
number DE 623/6-2.

\section*{References}
\bibliographystyle{iopart-num}
\bibliography{./my}

\providecommand{\newblock}{}
\begin{thebibliography}{10}
\expandafter\ifx\csname url\endcsname\relax
  \def\url#1{{\tt #1}}\fi
\expandafter\ifx\csname urlprefix\endcsname\relax\def\urlprefix{URL }\fi
\providecommand{\eprint}[2][]{\url{#2}}

\bibitem{Amoroso:2022eow}
Amoroso S {\em et~al.\/} 2022 {\em Acta Phys. Polon. B\/} {\bf 53} A1
  (\textit{Preprint} \eprint{2203.13923})

\bibitem{Ball:2021leu}
Ball R~D {\em et~al.\/} (NNPDF) 2022 {\em Eur. Phys. J. C\/} {\bf 82} 428
  (\textit{Preprint} \eprint{2109.02653})

\bibitem{pineline}
Barontini A, Candido A, Cruz-Martinez J~M, Hekhorn F and Schwan C 2023
  (\textit{Preprint} \eprint{2302.12124})

\bibitem{Gao:2017yyd}
Gao J, Harland-Lang L and Rojo J 2018 {\em Phys. Rept.\/} {\bf 742} 1--121
  (\textit{Preprint} \eprint{1709.04922})

\bibitem{Accardi:2012qut}
Accardi A {\em et~al.\/} 2016 {\em Eur. Phys. J. A\/} {\bf 52} 268
  (\textit{Preprint} \eprint{1212.1701})

\bibitem{Anderle:2021wcy}
Anderle D~P {\em et~al.\/} 2021 {\em Front. Phys. (Beijing)\/} {\bf 16} 64701
  (\textit{Preprint} \eprint{2102.09222})

\bibitem{NNPDF:2021uiq}
Ball R~D {\em et~al.\/} (NNPDF) 2021 {\em Eur. Phys. J. C\/} {\bf 81} 958
  (\textit{Preprint} \eprint{2109.02671})

\bibitem{Carrazza:2020gss}
Carrazza S, Nocera E~R, Schwan C and Zaro M 2020 {\em JHEP\/} {\bf 12} 108
  (\textit{Preprint} \eprint{2008.12789})

\bibitem{christopher_schwan_2023_7499507}
Schwan C, Candido A, Hekhorn F and Carrazza S 2023 Nnpdf/pineappl: v0.5.9
  \urlprefix\url{https://doi.org/10.5281/zenodo.7499507}

\bibitem{NNPDF:2014otw}
Ball R~D {\em et~al.\/} (NNPDF) 2015 {\em JHEP\/} {\bf 04} 040
  (\textit{Preprint} \eprint{1410.8849})

\bibitem{candido_alessandro_2022_6285149}
Candido A, Hekhorn F and Magni G 2022 N3pdf/yadism: Fonll-b
  \urlprefix\url{https://doi.org/10.5281/zenodo.6285149}

\bibitem{candido_alessandro_2022_6340153}
Candido A, Hekhorn F and Magni G 2022 N3pdf/eko: Paper
  \urlprefix\url{https://doi.org/10.5281/zenodo.6340153}

\bibitem{Candido:2022tld}
Candido A, Hekhorn F and Magni G 2022 {\em Eur. Phys. J. C\/} {\bf 82} 976
  (\textit{Preprint} \eprint{2202.02338})

\bibitem{AbdulKhalek:2022laj}
Abdul~Khalek R, Bertone V, Khoudli A and Nocera E~R 2022 {\em Phys. Lett. B\/}
  {\bf 834} 137456 (\textit{Preprint} \eprint{2204.10331})

\bibitem{Ball:2022qtp}
Ball R~D, Candido A, Forte S, Hekhorn F, Nocera E~R, Rojo J and Schwan C 2022
  {\em Eur. Phys. J. C\/} {\bf 82} 1160 (\textit{Preprint} \eprint{2209.08115})

\bibitem{Dokshitzer:1977sg}
Dokshitzer Y~L 1977 {\em Sov. Phys. JETP\/} {\bf 46} 641--653 [Zh. Eksp. Teor.
  Fiz.73,1216(1977)]

\bibitem{Gribov:1972ri}
Gribov V~N and Lipatov L~N 1972 {\em Sov. J. Nucl. Phys.\/} {\bf 15} 438--450
  [Yad. Fiz.15,781(1972)]

\bibitem{Altarelli:1977zs}
Altarelli G and Parisi G 1977 {\em Nucl. Phys.\/} {\bf B126} 298--318

\bibitem{Bertone:2013vaa}
Bertone V, Carrazza S and Rojo J 2014 {\em Comput. Phys. Commun.\/} {\bf 185}
  1647--1668 (\textit{Preprint} \eprint{1310.1394})

\bibitem{Botje:2010ay}
Botje M 2011 {\em Comput.Phys.Commun.\/} {\bf 182} 490--532 (\textit{Preprint}
  \eprint{1005.1481})

\bibitem{barontini_andrea_2023_7636142}
Barontini A, Candido A, Hekhorn F and Magni G 2023 N3pdf/banana: polarized toy
  \urlprefix\url{https://doi.org/10.5281/zenodo.7636142}

\bibitem{PER-GRA:2007}
P\'erez F and Granger B~E 2007 {\em Computing in Science and Engineering\/}
  {\bf 9} 21--29 ISSN 1521-9615 \urlprefix\url{https://ipython.org}

\bibitem{Forte:2010ta}
Forte S, Laenen E, Nason P and Rojo J 2010 {\em Nucl. Phys. B\/} {\bf 834}
  116--162 (\textit{Preprint} \eprint{1001.2312})

\bibitem{Giele:2002hx}
Giele W {\em et~al.\/} 2002 {The QCD / SM working group: Summary report} {\em
  {2nd Les Houches Workshop on Physics at TeV Colliders}\/} pp 275--426
  (\textit{Preprint} \eprint{hep-ph/0204316})

\bibitem{Dittmar:2005ed}
Dittmar M {\em et~al.\/} 2005  (\textit{Preprint} \eprint{hep-ph/0511119})

\bibitem{Vogt:2004ns}
Vogt A 2005 {\em Comput. Phys. Commun.\/} {\bf 170} 65--92 (\textit{Preprint}
  \eprint{hep-ph/0408244})

\bibitem{Buckley:2014ana}
Buckley A, Ferrando J, Lloyd S, Nordstr\"om K, Page B, R\"ufenacht M,
  Sch\"onherr M and Watt G 2015 {\em Eur. Phys. J. C\/} {\bf 75} 132
  (\textit{Preprint} \eprint{1412.7420})

\bibitem{Xie:2021equ}
Xie K, Hobbs T~J, Hou T~J, Schmidt C, Yan M and Yuan C~P (CTEQ-TEA) 2022 {\em
  Phys. Rev. D\/} {\bf 105} 054006 (\textit{Preprint} \eprint{2106.10299})

\bibitem{Cridge:2021pxm}
Cridge T, Harland-Lang L~A, Martin A~D and Thorne R~S 2022 {\em Eur. Phys. J.
  C\/} {\bf 82} 90 (\textit{Preprint} \eprint{2111.05357})

\bibitem{Bertone:2017bme}
Bertone V, Carrazza S, Hartland N~P and Rojo J (NNPDF) 2018 {\em SciPost
  Phys.\/} {\bf 5} 008 (\textit{Preprint} \eprint{1712.07053})

\bibitem{NNPDF:2019ubu}
Abdul~Khalek R {\em et~al.\/} (NNPDF) 2019 {\em Eur. Phys. J. C\/} {\bf 79} 931
  (\textit{Preprint} \eprint{1906.10698})

\bibitem{McGowan:2022nag}
McGowan J, Cridge T, Harland-Lang L~A and Thorne R~S 2023 {\em Eur. Phys. J.
  C\/} {\bf 83} 185 (\textit{Preprint} \eprint{2207.04739})

\bibitem{Caola:2022ayt}
Caola F, Chen W, Duhr C, Liu X, Mistlberger B, Petriello F, Vita G and
  Weinzierl S 2022 {The Path forward to N$^3$LO} {\em {2022 Snowmass Summer
  Study}\/} (\textit{Preprint} \eprint{2203.06730})

\end{thebibliography}

\end{document}